\documentclass[12pt]{article}
\usepackage{amsmath,amssymb}

\begin{document}
\begin{titlepage}

\begin{flushright}
hep-th/0703056
\end{flushright}
\vspace{1cm}

\begin{center}
{\Large\bf Nonlinear Realizations in Tensorial Superspaces \\[8pt]
and Higher Spins}
\vspace{1.5cm}

{\large\bf
Evgeny Ivanov}
\vspace{1cm}

{\it Bogoliubov Laboratory of Theoretical Physics,
JINR, \\
141980, Dubna, Moscow Region, Russia}\\
{\tt eivanov@theor.jinr.ru}\\[8pt]
\end{center}
\vspace{2cm}

\begin{abstract}
\noindent
In the first part of the talk I report on surprising
relations between higher spin (HS) theory
and nonlinear realizations of the supergroup $OSp(1|8)$, a minimal
superconformal extension of $N=1$, $4D$ supersymmetry with tensorial
charges. The second part is a review of the ``master'' model
of HS particle which makes manifest the classical
equivalence of two previously known HS particle formulations
and gives rise to new massless HS multiplets.
\end{abstract}
\vspace{6cm}

\begin{center}
{\it Talk given at the XXII Max Born Symposium ``Quantum, Super and Twistors''
in honor of the 70th birthday of Prof. Jerzy Lukierski, Wroclaw, 26 - 29 September 2006}
\end{center}

\end{titlepage}

\setcounter{page}{1}
\noindent{\bf 1. Introduction.} Since the seminal papers by Fradkin and Vasiliev \cite{FV}, the
theory of higher spin (HS) fields attracts vast interest (see
e.g. \cite{rev,rev2} and refs. therein), especially due to  its profound
relations to string theory. A concise and suggestive way of dealing with higher
spins is to treat them in spaces with additional
coordinates, e.g. the tensorial ones \cite{ill4}--\cite{ill6}.

M. Vasiliev has
shown in \cite{ill5} that the free $4D$ HS field theory can
be described by the pair of bosonic and fermionic fields $b(Y),
f_{\hat{\alpha}} (Y)$ ($\hat{\alpha} = 1,2,3,4$) defined on
ten-dimensional real tensorial space
  \begin{equation}\label{ll1}
    Y^{\hat\alpha \hat\beta}=
    Y^{\hat\beta\hat\alpha} = \frac{1}{2}\
    x^{m}(\gamma_m)^{\hat\alpha \hat\beta} +
    \frac{1}{4}\ y^{[m \, n]}
    (\gamma_{[m\, n]})^{\hat\alpha\hat\beta} \,,
\end{equation}
where $x^m$ are Minkowski space coordinates. A nice superfield form
of these equations, in the tensorial superspace $R^{(10|4)} =
(Y^{\hat\alpha\hat\beta} , \theta^{\hat\alpha})$, was given in \cite{ill6}.

One of the goals of my talk is to argue \cite{ILu}
that an adequate setting for the HS equations in extended (super)spaces
is provided by nonlinear realizations of the supergroup
$OSp(1|8)$, a minimal superconformal extension of
$N=1,\, 4D$ supersymmetry with tensorial charges \cite{ill8}-\cite{fpn}.
Besides reproducing the HS equations of refs. \cite{ill5,ill6}
in a nice geometric way, such framework opens new avenues in the
HS theory. In particular, it suggests an important role of the $OSp(1|8)$ generalization
of the $N=1,\, 4D$ chirality concept which underlies
gauge $N=1,\, 4D$ theories including supergravity \cite{OS}.

The simple and, at the same
time, powerful device for the analysis of the geometric structure
of extended (super)spaces is provided by (super)particles propagating
in them. As the second subject of my talk I describe the new ``master'' model
of the bosonic HS particle \cite{FI-CQG} which encompasses two previously known
HS particle models and gives rise to new massless free HS multiplets.
\vspace{0.3cm}

\noindent{\bf 2. $OSp(1|8)$ as a generalized superconformal group.}
The even (bosonic) sector of the superalgebra $osp(1|8)$ is
the generalized $4D$ conformal algebra
$sp(8)$ which is a closure of
the standard conformal algebra $so(2,4)$ and the algebra $sl(4,R)$.
The algebra $so(2,4)\simeq su(2,2)$ is spanned by the generators
($L_{\alpha \dot{\beta}}, {\overline{L}}_{\dot{\alpha}\dot{\beta}},
P_{\alpha \dot{\beta}}, K_{\alpha\dot{\beta}}, D$). The algebra
$sl(4,R)$ is spanned by the generators $(L_{\alpha
\beta}, \overline{L}_{\dot{\alpha}\dot{\beta}}, A, F_{\alpha
\dot{\beta}}, \overline{F}_{\alpha\dot{\beta}})$. The extra
generators $A$, $F_{\rho \dot{\tau}}$,
$\overline{F}_{\rho\dot{\tau}}\equiv (F_{\tau\dot\rho})^*\, $
satisfy the following non-zero commutation relations
\begin{eqnarray} \label{ll2.1a}  [
F_{\alpha\dot\beta}, \overline{F}_{\beta\dot\nu} ]
 = 2 \epsilon_{\alpha\beta}\epsilon_{\dot\beta\dot\nu} A
+ 2 \left(\epsilon_{\alpha\beta}\bar L_{\dot\beta\dot\nu}
- \epsilon_{\dot\beta\dot\nu} L_{\alpha\beta}\right), \; [ A, F_{\alpha\dot\beta}] = 2\,
F_{\alpha\dot\beta}
\end{eqnarray}
(and c.c.). The algebra $sp(8)$ includes the following additional 12 Abelian generators:
($Z_{\alpha {\beta}}, \overline{Z}_{\dot{\alpha}\dot{\beta}}$)
describing six standard tensorial translations and (${\widetilde{Z}}_{\alpha \beta},
\overline{{\widetilde{Z}}}_{\dot{\alpha}\dot{\beta}}$) describing six
 conformal tensorial translations.

The odd (fermionic) sector of $osp(1|8)$ involves $N=1$ super
 Poincar\'e generators $Q_\alpha, \bar Q_{\dot\alpha}$
and the generators $S_\alpha, \bar S_{\dot\alpha}$ of
 conformal supersymmetry:
\begin{eqnarray} \label{llnew2.13a}
&&\{Q_\alpha, \bar Q_{\dot\alpha}\} = 2 P_{\alpha\dot\alpha}\,,
 \quad
\{Q_\alpha, Q_{\beta}\} = 2 Z_{\alpha\beta}\,, \nonumber \\
&&
\{S_\alpha, \bar S_{\dot\alpha}\} = 2 K_{\alpha\dot\alpha}\,,
\quad
\{S_\alpha, S_{\beta}\} = 2 {\widetilde{Z}}_{\alpha\beta}\,, \nonumber \\
&&
\{Q_\alpha, \bar S_{\dot\beta} \} = F_{\alpha\dot\beta}\,, \quad
\{Q_\alpha, S_\beta \} = \epsilon_{\alpha\beta}\left(iD -\frac 12 A \right)
+ L_{\alpha\beta}
\end{eqnarray}
(and c.c.). The remaining $OSp(1|8)$ (anti)commutators can be found in \cite{ILu}.
\vspace{0.3cm}

\noindent{\bf 3. Nonlinear realizations of $OSp(1|8)$.}
We construct nonlinear realization of $OSp(1|8)$ in the
supercoset $OSp(1|8)/SL(4,R)$.
It contains generators
$\{Q, \bar Q, P, Z, \overline{Z}$, $S, \bar S, K, \tilde{Z}, \overline{\tilde{Z}}, D\}$
(where, for brevity, we suppressed Lorentz indices) and is parametrized
by the coordinates
\begin{equation}
{\widetilde{Y}}^M \equiv (Y^{\hat{\alpha}\hat{\beta}}, \theta^\alpha,
\bar\theta^{\dot\alpha})\,, \;Y^{\hat{\alpha}\hat{\beta}} \equiv
(x^{\alpha\dot\beta}, z^{\alpha\beta}, \bar{z}^{\dot\alpha\dot\beta}),\label{TSup}
\end{equation}
the fermionic coordinates $\theta, \bar\theta$ being associated with
the generators $Q, \bar Q$ and the remaining bosonic coordinates with
$P, Z, \overline{Z}$. In addition, it involves the Goldstone superfields
$\{\psi^\alpha(\widetilde{Y}),
\bar\psi^{\dot\alpha}(\widetilde{Y}), k_{\alpha\dot\alpha}(\widetilde{Y}),
t_{\alpha\beta}(\widetilde{Y}), \bar{t}_{\dot\alpha\dot\beta}(\widetilde{Y}),
\phi(\widetilde{Y})\}$ associated with the generators $S, \bar S, K, \tilde{Z},
\overline{\tilde{Z}}$ and $D$, respectively.

For the  supercoset elements we use the exponential parametrization
\begin{equation}
 G = e^{i(\theta Q
+ \bar\theta \bar Q)}\,e^{i(x\cdot P + z\cdot {Z})}\,e^{i\phi D} \,e^{i(k\cdot K + t\cdot
{\widetilde{Z}})}\, e^{i(\psi S + \bar\psi \bar S)}\,.\label{suparam}
\end{equation}
The left-covariant Cartan
forms are defined by:
\begin{eqnarray} G^{-1}dG &=& i (\,
\Omega_Q\cdot Q + \Omega_S\cdot S + \Omega_P\cdot P + \Omega_Z\cdot
Z + \Omega_D D + \Omega_K\cdot
K
\nonumber \\ &&  + \,
\Omega_{\widetilde{Z}}{}\cdot {\widetilde{Z}} + \Omega_L\cdot L +
\Omega_A A + \Omega_F\cdot F \,) \equiv i \Omega\,. \label{Supforms}
 \end{eqnarray}
All the Goldstone superfields can be covariantly expressed through the
dilatonic one $\phi(\tilde{Y})$ by the inverse Higgs \cite{ill12} constraint
 \begin{eqnarray}
&& \Omega_D = 0 \quad \Rightarrow \label{IHconstr} \\
&& k_{\alpha\dot\alpha} = -
e^{-\phi}\,\partial_{\alpha\dot\alpha}\phi\,, \;\;t_{\alpha\beta} =
\frac 12 e^{-\phi}\,\partial_{\alpha\beta}\phi\,, \;\; \bar
t_{\dot\alpha\dot\beta} = \frac 12
e^{-\phi}\,\partial_{\dot\alpha\dot\beta}\phi \,, \label{IH}\\
&&\psi_\alpha = -e^{-\frac
12 \phi}\,D_\alpha\phi\,, \quad \bar\psi_{\dot\alpha} = -e^{-\frac
12 \phi}\,D_{\dot\alpha}\phi\,, \label{IH2} \\
&& D_\alpha = \frac{\partial}{\partial \theta^\alpha} -
i\bar\theta^{\dot\beta}\partial_{\alpha\dot\beta} +i
\theta^\beta\partial_{\alpha\beta}\,,\;\bar D_{\dot\alpha} =
-\frac{\partial}{\partial \bar\theta^{\dot\alpha}} +
i\theta^{\beta}\partial_{\beta\dot\alpha} - i
\bar\theta^{\dot\beta}\partial_{\dot\alpha\dot\beta}\,.
\label{Ddefin}
\end{eqnarray}

The HS field dynamics in this setting also arises as a sort of dynamical
inverse Higgs effect. Namely, it amounts to the vanishing of the full set
of the $OSp(1|8)$ covariant spinor derivatives of the Goldstone superfields
$\psi^\alpha, \bar\psi^{\dot\alpha}$ (defined as the projections of the
Cartan forms $\Omega_S^\alpha, \bar\Omega^{\dot\alpha}_S, \Omega_D$ on the
$\theta, \bar\theta$ covariant differentials $\Omega^\beta_Q, \bar\Omega^{\dot\beta}_Q$):
\begin{eqnarray}
\nabla_\beta\psi^\alpha = 0\,, \;\;
\bar\nabla_{\dot\beta}\psi^\alpha = 0\,, \;\;\nabla_\beta\phi = 0 \quad \mbox{and c.c.}\,.
\label{Eqcova}
\end{eqnarray}
These equations yield both the expressions (\ref{IH}), (\ref{IH2})
and the equations
\begin{equation}
(D)^2e^{\phi} = (\bar D)^2 e^{\phi}=0\,, \;\;
\left[D^\alpha, \bar D^{\dot\alpha}\right] e^{\phi} = 0\,.
\label{EQ}
\end{equation}
Eqs. (\ref{EQ}) are recognized as the two-component spinor form of the equation suggested in
\cite{ill6} (for $\Phi = e^\phi$).
\vspace{0.3cm}

\noindent{4. \bf Tensorial chiral superspace.}
The underlying $N=1$ supergravity gauge group  is a group of general
diffeomorphisms of chiral $N=1$ superspace $C^{(4|2)} =
(x_L^m, \theta^\alpha_L)$ \cite{OS}. This manifests the
fundamental role of the principle of preserving chiral
representations in $N=1$ supergravity.
The question arises whether an analog of this principle exists for
  higher-spin generalization of $N=1$ supergravity.
From the analysis of the full set of the (anti)commutation relations
of the superalgebra $osp(1|8)$ it follows that the minimal analog of
$C^{(4|2)}$ is the coset spanned by the following generators
$(P_{\alpha\dot\alpha}, Z_{\alpha\beta}, F_{\beta\dot\beta},
Q_\alpha )$, i.e. it contains only one
holomorphic half of the tensorial central charges and, in addition,
the complex generator $F_{\beta\dot\beta}\,$.
Thus the set of the relevant coset parameters, i.e. $C^{(11|2)} =
(x_L^{\alpha\dot\beta}, z_L^{\alpha\beta}, f_L^{\alpha\dot\beta},
\theta^\alpha_L) \equiv (\,Y_L\,)$ is closed
under the left action of the supergroup $OSp(1|8)$ and provides a
natural generalization of $C^{(4|2)}$.
It is interesting to inquire whether some higher-spin
 dynamics can be associated with superfields given on $C^{(11|2)}$ as an alternative
 to eqs. (\ref{EQ}) and what is the theory enjoying invariance under general
 diffeomorphisms of $C^{(11|2)}$ (the HS analog of $N=1, 4D$ conformal
 supergravity?) Some examples of
 the $OSp(1|8)$ invariant actions of the tensorial chiral superfields were
 presented in \cite{ILu}. It still remains to reveal their component field contents and
 their relation to the dynamics of HS fields.
\vspace{0.3cm}

\noindent{\bf 5. HS particles.}
The unfolded formulation of the HS theory \cite{ill5} is
reproduced by quantizing the tensorial
particle~\cite{ill3}, \cite{ill6}, or the equivalent HS
particle~\cite{ill5}, in which tensorial coordinates were gauged
away.
There also exists a
different formulation of the HS particle exhibiting invariance
under the even counterpart of supersymmetry~\cite{FedLuk}.

In the unfolded formulation without tensorial coordinates \cite{ill5}
the basic equation for the real HS field $\Phi(x,
y,\bar y)$ reads
\begin{equation}\label{unfold-eq}
\left(\partial_{\alpha\dot\alpha} +i\frac{\partial}{\partial y^{\alpha}}
\frac{\partial}{\partial \bar y^{\dot\alpha}} \right) \Phi = 0\,,
\end{equation}
where $y^{\alpha}$ is a commuting Weyl spinor,
$\bar y^{\dot\alpha}=\overline{(y^{\alpha})}$.
Solution of eq.~(\ref{unfold-eq}) can be found,
assuming the polynomial dependence of $\Phi$ on $y^{\alpha}$, $\bar y^{\dot\alpha}$
\begin{equation}\label{wf-tens}
\Phi(x, y, \bar y)  =\sum_{m=0}^{\infty}\sum_{n=0}^{\infty} y^{\alpha_1}\ldots y^{\alpha_m}
\bar y^{\dot\alpha_1}\ldots\bar y^{\dot\alpha_n} \varphi_{\alpha_1 \ldots \alpha_m
\dot\alpha_1 \ldots \dot\alpha_n}(x)\,.
\end{equation}
The independent space-time fields in this expansion are self--dual $\varphi_{\alpha_1 \ldots
\alpha_m}$ and anti--self--dual $\varphi_{\dot\alpha_1 \ldots
\dot\alpha_n}$ field strengths of all helicities. Eq.~(\ref{unfold-eq})
leads to Klein--Gordon and Dirac equations for them.

A classical counterpart of this unfolded formulation is the particle action
\begin{equation}\label{act-1}
S_{1}=\int d\tau \left( \lambda_\alpha \bar\lambda_{\dot\alpha} \dot x^{\dot\alpha\alpha} +
\lambda_\alpha\dot y^\alpha + \bar\lambda_{\dot\alpha}\dot{\bar y}^{\dot\alpha} \right).
\end{equation}
The spinors $\lambda_\alpha$, $\bar\lambda_{\dot\alpha}$ are
canonical momenta for $y^{\alpha}$, $\bar y^{\dot\alpha}\,$. The
constraints
\begin{equation}\label{P-res}
P_{\alpha\dot\alpha} - \lambda_\alpha
\bar\lambda_{\dot\alpha} \approx 0
\end{equation}
are first class and after quantization reproduce the unfolded equation~(\ref{unfold-eq}).

A different model of the massless HS particle was
proposed in~\cite{FedLuk}:
\begin{equation}\label{act-bsusy}
S_{2}=\int d\tau \left( P_{\alpha\dot\alpha} \omega^{\dot\alpha\alpha} - e
P_{\alpha\dot\alpha}P^{\alpha\dot\alpha} \right),\;
\omega^{\dot\alpha\alpha} \equiv \dot x^{\dot\alpha\alpha} -i
{\bar\zeta}{}^{\dot\alpha}\dot{\zeta}{}^{\alpha}+i
\dot{\bar\zeta}{}^{\dot\alpha}{\zeta}^{\alpha}\,.
\end{equation}
The crucial difference of (\ref{act-bsusy}) from the $N=1$ superparticle action is
that the Weyl spinor $\zeta^\alpha$, $\bar\zeta^{\dot\alpha}=
(\overline{\zeta^\alpha})$, is commuting. This model is manifestly invariant under the even
counterpart of $4D$ supersymmetry~\cite{FedZim,Lecht,FedLuk,FIL}
\begin{equation}\label{susy}
\delta x^{\dot\alpha\alpha} =
i(\bar\epsilon^{\dot\alpha}\zeta^\alpha
-\bar\zeta^{\dot\alpha}\epsilon^\alpha) \, ,\quad \delta
\zeta^\alpha = \epsilon^\alpha \, ,\quad \delta
\bar\zeta^{\dot\alpha}= \bar\epsilon^{\dot\alpha}\,,
\end{equation}
where $\epsilon^\alpha$ is a commuting Weyl spinor.

The set of the Hamiltonian constraints of the system includes the
mass-shell constraint $
P_{\alpha\dot\alpha}P^{\alpha\dot\alpha}\approx 0 $ and the even
spinor constraints
\begin{equation}\label{cons-D}
D_\alpha\equiv \pi_\alpha
+iP_{\alpha\dot\alpha}\bar\zeta^{\dot\alpha}\approx
0\,,\qquad\qquad \bar D_{\dot\alpha}\equiv \bar\pi_{\dot\alpha}
-i\zeta^{\alpha}P_{\alpha\dot\alpha}\approx 0\,,
\end{equation}
where $\pi_\alpha$ and $\bar \pi_{\dot\alpha}$ are conjugate momenta
for $\zeta^\alpha$ and $\bar \zeta^{\dot\alpha}$.

The wave function of the HS particle model~(\ref{act-bsusy}) is
\begin{equation}\label{even-sfield}
\Psi(x_{\!\scriptscriptstyle L}, \zeta)=
\sum_{n=0}^{\infty} \zeta^{\alpha_1}\ldots \zeta^{\alpha_n}
\psi_{\alpha_1 \ldots \alpha_n}
(x_{\!\scriptscriptstyle L})\,,\quad \bar D_{\dot\alpha} \Psi =0\,,
\end{equation}
where $x_{\!\scriptscriptstyle L}^{\dot\alpha\alpha} =
x^{\dot\alpha\alpha} +i
{\bar\zeta}{}^{\dot\alpha}{\zeta}{}^{\alpha}\,$.
This field is subjected to the equations
\begin{equation}\label{1-cl-con}
\partial_{\!\scriptscriptstyle L}^{\dot\alpha\alpha} \partial_{\alpha} \, \Psi =
0\,, \qquad
\partial_{\!\scriptscriptstyle L}^{\dot\alpha\alpha}
\partial_{{\!\scriptscriptstyle L}\,\alpha\dot\alpha}\, \Psi = 0\,,
\end{equation}
which are quantum counterparts of the first class constraints. Due to
eqs.~(\ref{1-cl-con}) the fields in the
expansion~(\ref{even-sfield}) are complex self--dual field
strengths of the massless particles of all helicities. As a
result, the spectrum contains all helicities, every non-zero
helicity appearing only once. In this picture the scalar field is
complex, as opposed to the unfolded formulation.
\vspace{0.3cm}

\noindent{\bf 6. Master HS particle model.}
The master HS system action \cite{FI-CQG} involves the variables of both
systems~(\ref{act-1}) and
(\ref{act-bsusy}) plus a complex scalar $\eta$:
\begin{eqnarray}\label{act-mast}
S =\int d\tau [\lambda_\alpha \bar\lambda_{\dot\alpha} \omega^{\dot\alpha\alpha} +
\lambda_\alpha\dot y^\alpha + \bar\lambda_{\dot\alpha} \dot{\bar y}^{\dot\alpha}
+i(\eta\dot{\bar\eta} - \dot\eta\bar\eta)
+2i(\eta \bar\lambda_{\dot\alpha} \dot{\bar\zeta}{}^{\dot\alpha} -
\bar\eta\dot{\zeta}{}^{\alpha}\lambda_\alpha)
- l({\cal N} - c)].
\end{eqnarray}
The field $l$ acts as a Lagrange multiplier for the
constraint
\begin{equation}\label{H-mast}
{\cal N} -c \equiv i\,(y^\alpha\lambda_\alpha
-\bar\lambda_{\dot\alpha} {\bar y}^{\dot\alpha})- 2\eta\bar\eta -c
\approx 0\,,
\end{equation}
which generates, in the Hamiltonian formalism, local $U(1)$
transformations of the involved complex fields (except for $\zeta$, $\bar\zeta$).

The action~(\ref{act-mast}) produces the following primary constraints
\begin{equation}\label{T-mast}
T_{\alpha\dot\alpha}\equiv P_{\alpha\dot\alpha} -\lambda_\alpha
\bar\lambda_{\dot\alpha}\approx 0\,,
\end{equation}
\begin{equation}\label{D-mast}
{\cal D}_\alpha\equiv D_\alpha +
2i\bar\eta\lambda_\alpha\approx 0\,,\qquad\qquad \bar{\cal D}_{\dot\alpha}\equiv
\bar{D}_{\dot\alpha} - 2i\eta\bar\lambda_{\dot\alpha}\approx 0
\end{equation}
and
\begin{equation}\label{g-mast}
g \equiv p_\eta +i\bar\eta \approx 0\,,\qquad \bar g \equiv \bar p_{\eta} -i\eta
\approx 0\,,
\end{equation}
where
$\lambda_\alpha$ and $\bar\lambda_{\dot\alpha}$ are treated as conjugate momenta
for $y^\alpha$ and
$\bar y^{\dot\alpha}\,$.
The constraints~(\ref{g-mast}) are second class and so can be treated in the strong sense by
introducing Dirac brackets.  Then $\eta, \bar\eta$ form the canonical pair:
$[\eta, \bar\eta ]_{{}_D}={\textstyle\frac{i}{2}}\,$.

The systems~(\ref{act-bsusy}) and (\ref{act-mast}) are (classically) equivalent
to each other in the common sector of their phase space which is singled out
by choosing the
definite sign of the energy $P_0$.
Two second class constraints contained in the spinor
constraints~(\ref{cons-D}) can be converted into the first class ones
by adding two degrees of freedom carried by the complex scalar field
$\eta$ and introducing a commuting spinor $\lambda_\alpha$ to ensure the Lorentz
covariance of
the new spinor constraints~(\ref{D-mast}).
The closure of the algebra of the new spinor constraints,
\begin{equation}\label{PB-con}
[{\cal D}_\alpha , \bar{\cal D}_{\dot\alpha} ]_{{}_D}=2i \, T_{\alpha\dot\alpha}\,,
\end{equation}
gives just the constraint~(\ref{T-mast}) resolving four-momentum in terms of the
spinor product.
This resolution is defined up to an arbitrary phase transformation of $\lambda_\alpha$
$\bar\lambda_{\dot\alpha}$. To ensure this $U(1)$ gauge invariance
in the full modified action, we just add the constraint~(\ref{H-mast}).

The world-line particle model~(\ref{act-1}) also follows from the
master model~(\ref{act-mast}) under a particular gauge choice. The
spinor constraints~(\ref{D-mast}) and the gauge-fixing condition
$\zeta^{\alpha} \approx 0$ together with its complex conjugate can
be used to eliminate the variables $\zeta^{\alpha}$, $\pi_\alpha$
and their complex conjugates. Then the constraint (\ref{H-mast}),
together with the gauge fixing condition $\chi \equiv \varphi - const \approx 0\,$,
can be used to gauge away the variable $\eta \equiv \sqrt{\rho}e^{i\varphi}$.
\vspace{0.3cm}

\noindent{\bf 7. First-quantized theory.}
The equations for the wave
function $F^{(q)} (x, \zeta, \bar\zeta,$ $y, \bar y, \eta)$ read
\begin{equation}\label{T-eq}
\left(\partial_{\alpha\dot\beta} +i\frac{\partial}{\partial
y^{\alpha}}
\frac{\partial}{\partial \bar y ^{\dot\beta}} \right) F^{(q)} = 0\,,
\end{equation}
\begin{equation}\label{D-eq}
\mbox{(a)}\;\;\left(D_{\alpha}+ \frac{\partial}{\partial \eta}
\frac{\partial}{\partial y^{\alpha}} \right) F^{(q)} = 0\,,
\quad\qquad \mbox{(b)}\;\;\left(\bar D_{\dot\alpha}- 2\eta
\frac{\partial}{\partial \bar y^{\dot\alpha}} \right)\, F^{(q)} =
0\,,
\end{equation}
\begin{equation}\label{H-eq}
\left(y^{\alpha}\frac{\partial}{\partial
y^{\alpha}} - \bar y^{\dot\alpha}\frac{\partial}{\partial
\bar y^{\dot\alpha}} - \eta\frac{\partial}{\partial
\eta} \right) F^{(q)} = q\,F^{(q)}\,.
\end{equation}
Here, the operators $ D_\alpha= -i (\partial_\alpha
+i\partial_{\alpha\dot\alpha} \bar\zeta^{\dot\alpha} )$ and ${\bar
D}_{\dot\alpha}= -i (\bar\partial_{\dot\alpha}
-i\zeta^{\alpha}\partial_{\alpha\dot\alpha})$ are quantum
counterparts of the ``covariant momenta''~(\ref{cons-D}).
The external $U(1)$ charge $q$ defined by~(\ref{H-eq}) is the quantum counterpart of
the constant $c$ in~(\ref{H-mast}).
Eq.~(\ref{H-eq}) implies $U(1)$
covariance of the wave function
\begin{equation}\label{u1-inv}
F^{(q)} (x, \zeta, \bar\zeta, e^{i\varphi} y,
e^{-i\varphi} \bar y, e^{-i\varphi} \eta) =
e^{qi\varphi} F^{(q)} (x, \zeta, \bar\zeta, y, \bar y, \eta)\,.
\end{equation}
Requiring $F^{(q)}$ to be single-valued restricts $q$ to the
integer values.

The set (\ref{T-eq})--(\ref{H-eq}) can be solved in two different ways. One can first
fix the dependence on the variables
$\zeta, \bar\zeta, \eta$ and end up with some complex function of $(x, y, \bar y)$ having
a fixed $U(1)$ charge $q$ and subjected to some holomorphic version of the unfolded
equation (\ref{unfold-eq}). Another way is to start by fixing the dependence on $y, \bar y, \eta$,
that gives rise to a set of equations like (\ref{1-cl-con}) for some chiral functions
of $(x, \zeta, \bar\zeta)$. Thus two alternative
formulations of the massless HS equations are naturally realized as two alternative
ways of solving the master set (\ref{T-eq})--(\ref{H-eq}). The eventual
sets of fields are the same in both cases. Since the wave
function $F^{(q)}$ carries an external $U(1)$ charge $q$ and bears a dependence
on the extra coordinate $\eta$, the set (\ref{T-eq})--(\ref{H-eq}) yields a richer
set of massless HS multiplets as compared to (\ref{unfold-eq}) or (\ref{1-cl-con}). Without
entering into details, let us list these multiplets, following ref. \cite{FI-CQG}.

$\underline{q = 0}$. The
space of physical states is spanned by the complex
self--dual field strengths $\phi_{\alpha_1 \ldots \alpha_k}$,
$k=0,1,2, \ldots\,$, of the massless particles of all integer and
half-integer helicities. Thus the case of $q=0$ basically amounts
to the standard HS multiplet of ref. \cite{ill5} (modulo the complexity of the scalar
field).

$\underline{q > 0}$. The space of physical states is spanned by the
self--dual field strengths of the massless particles with
helicities $\frac{q}{2},\frac{q}{2}+\frac{1}{2},
\frac{q}{2}+1,\ldots\,$. Thus the scalar field is
absent in the spectrum for non-zero positive $q\,$.

$\underline{q < 0}$. The physical fields describe
massless particles with all positive helicities including the zero one,
and a finite number of massless states with
negative helicities $-\frac{1}{2}, -1, \ldots, -\frac{|q|}{2}$.
Taken with its conjugate, this multiplet reveals a partial doubling of fields
with a given helicity.
\vspace{0.3cm}

\noindent{\bf 8. Summary.} To summarize, the new HS particle gives rise to an extended set of the
massless HS equations with novel HS multiplets as their solutions. Its
distinguishing features are the external $U(1)$ charge $q$
which basically coincides with the minimal helicity of given HS multiplet
and the presence of complex scalar coordinate $\eta$ on which the HS fields depend
holomorphically. It would be interesting to elaborate on a possible role of this
new coordinate in the geometry of HS (super)gravity (e.g. in the approach of ref.
\cite{ill6}). Also, it is tempting to incorporate supersymmetry into this picture
(see \cite{FIL} for the first steps) and to extend the master HS model to
include the tensorial coordinates, thus establishing links with the first half of
this talk.

\bigskip
\noindent {\bf Acknowledgments.} I would like to thank the Organizers of the
XXII Max Born Symposium for the kind hospitality in Wroclaw and offering me an opportunity
to give this talk and to express my great respect to my friend and co-author Jurek Lukierski
on occasion of his 70th Anniversary. I also thank my co-author Sergey Fedoruk.
My visit to Wroclaw was partially supported from the Bogoliubov-Infeld Program.  I acknowledge
a partial support from the RFBR grant 06-02-16684 and the grant INTAS-05-7928.

\end{document}